\documentclass[
   ,final            
  ]
  {aipproc}

\layoutstyle{6x9}
\usepackage{epsfig}

\begin{document}

\title{Milagro Search for Very High Energy Emission from Gamma-Ray Bursts in the \em{Swift} Era}
\author{P.~M.~Saz Parkinson for the Milagro Collaboration\footnote{
A.~Abdo,
B.~T.~Allen,
R.~Atkins,
D.~Berley,
E.~Blaufuss,
S.~Casanova,
D.~G.~Coyne,
B.~L.~Dingus,
R.~W.~Ellsworth,
L.~Fleysher,
R.~Fleysher,
M.~M.~Gonzalez,
J.~A.~Goodman,
E.~Hays,
C.~M.~Hoffman,
L.~A.~Kelley,
C.~P.~Lansdell,
J.~T.~Linnemann,
J.~E.~McEnery,
A.~I.~Mincer,
M.~F.~Morales,
P.~Nemethy,
D.~Noyes,
J.~M.~Ryan,
F.~W.~Samuelson,
P.~M.~Saz~Parkinson,
A.~Shoup,
G.~Sinnis,
A.~J.~Smith,
G.~W.~Sullivan,
V.~Vasileiou,
G.~P.~Walker,
D.~A.~Williams,
X.~W.~Xu
and 
G.~B.~Yodh}}{
  address={Santa Cruz Institute for Particle Physics, University of California, 
1156 High Street, Santa Cruz, CA 95064}
}

\begin{abstract}

The recently launched {\em Swift} satellite is providing an unprecedented number of rapid
and accurate Gamma-Ray Burst (GRB) localizations, facilitating a flurry of follow-up 
observations by a large number of telescopes at many different wavelengths. The Very
High Energy (VHE, $>$100 GeV) regime has so far been relatively unexplored. 
Milagro is a wide field of view (2 sr) and high duty cycle ($>$ 90\%) ground-based
gamma-ray telescope which employs a water Cherenkov detector to monitor the northern
sky almost continuously in the 100 GeV to 100 TeV energy range. We have searched the
Milagro data for emission from the most recent GRBs identified within our field of view.
These include three {\em Swift} bursts which also display late-time X-ray flares. We have searched 
for emission coincident with these flares. No significant detection was made. A 99\%
confidence upper limit is provided for each of the GRBs, as well as the flares.

\end{abstract}

\maketitle

\section{Introduction}

Some of the most important contributions to our understanding of gamma-ray bursts have come
from observations of afterglows over a wide spectral range~\citep{2000ARA&A..38..379V}. Very 
little, however, is known about the broadband spectra of GRBs in the prompt phase, due
to its short duration. Many GRB production models predict a fluence at TeV comparable to 
that at MeV scales~\citep{dermer00,pilla98,zhang01}. Almost all GRBs are detected in the energy
range between 20 keV and 1 MeV, though several have been observed above 100 MeV by EGRET, 
indicating that the spectrum of GRBs extends at least out to 1 GeV~\citep{dingus01}. A second
component was also found in one burst which extended up to at least 200 MeV and 
had a much slower temporal decay than the main burst~\citep{gonzalez03}. It is unclear how high 
in energy this component extends to and whether it is similar to the inverse Compton peak seen 
in many TeV sources. At very high energies there has been no conclusive emission detected for 
any single
GRB, though a search for counterparts to 54 BATSE bursts with Milagrito, a prototype of Milagro, 
found evidence for emission from one burst, with a significance slightly greater than 
3$\sigma$~\citep{atkins00a}. At these high energies, gamma rays suffer from attenuation due to the
extra-galactic background light (EBL), which is 
redshift-dependent~\citep{stecker98,primack99,primack05}, making GRBs above z$>$0.5
very difficult to observe. Here, we describe our search for VHE emission from 20 GRBs which
have occurred within the field of view of Milagro\footnote{We have included GRB 051103 in our sample, 
despite being at a relatively large zenith angle, given its potential interest and possible
proximity.} between December 2004 and December 2005. Recently, {\em Swift} has also observed bright 
X-ray flares from GRB afterglows, sometimes of comparable energy
to the burst itself~\citep{burrows05,falcone05}. Although the most significant of these flares (from GRB 050502B) 
was unfortunately outside the field of view of Milagro, six other flares from three different
GRBs were observable~\citep{falcone05b}. We present the upper limits derived from our observations.

\section{Data Analysis and Results}

A search for an excess of events above those due to the background was made for each of the 20 bursts
in our sample, listed in Table~\ref{grb_table}, as well as for the flares listed in Table~\ref{grb_table2}. The number of events falling within a 1.6 degree bin
was summed for the relevant duration (column 2 of Table~\ref{grb_table} or column 4 of Table~\ref{grb_table2}). An estimate of the 
number of background events was made by characterizing the angular distribution of the background
using two hours of data surrounding the burst, using a technique known as ``direct integration''
~\citep{atkins03b}. For those bursts whose redshift is known, we compute the effect of the 
absorption, according to the model of \cite{primack05} and print the upper limits in bold. 
No significant emission was detected. Our upper limits are given in column seven of Table~\ref{grb_table} (for the GRBs) and Table~\ref{grb_table2} (for the flares).

Some of the more interesting bursts in Table~\ref{grb_table} 
include GRB 050509b, the second \emph{short/hard} burst detected by \emph{Swift}, with a reported 
duration of 30 ms and a relatively low fluence of 2.3$\times10^{-8}$ erg cm$^{-2}$ in the 15--350 
keV range~\cite{barthelmy}. Although Milagro detected no emission from this burst~\cite{saz}, the very 
favorable zenith angle ($10^\circ$) and possible low redshift of 0.226 provide the opportunity
to set interesting upper limits for TeV emission from this burst. Another burst that deserves mention
is GRB 051103, a short burst detected by the IPN, possibly originating from the 
nearby ($<$ 4 Mpc) galaxy M81~\citep{051103}. Although this burst was not at a favorable zenith 
angle for Milagro ($\theta >$45$^\circ$), we chose to analyse the data, due 
to its potential interest. If this burst were associated with M81, as has been suggested, the 
almost complete absence of absorption would make the Milagro upper limit on the fluence 
about 20\% of the measured fluence in the X-ray band~\citep{saz2}. Finally, GRB 051109 occurred 
at a zenith angle of less than 10$^\circ$ for Milagro, making it the burst with the best 
location of our sample. Unfortunately, this burst had a measured redshift of 2.346, making the 
Milagro limit, once absorption is taken into account, several orders of magnitude larger than 
the measured fluence in the X-ray band~\citep{saz3}.

In Table~\ref{grb_table2} we list 6 flares detected by {\em Swift} from
3 GRBs which were in Milagro's field of view~\citep{falcone05}. Milagro detected no significant 
emission from any of these flares. The brightest of these flares, from GRB 050607, had a fluence in 
the X-ray range of roughly 1.5$\times 10^{-7}$ erg cm$^{-2}$~\citep{king05}, placing the Milagro 
limit about an order of magnitude higher.


\begin{table}
\begin{tabular}{llllllll}
\hline
	 \tablehead{1}{l}{b}{GRB} &
	 \tablehead{1}{l}{b}{T90/Dur.\tablenote{Duration of burst.}} & 
	 \tablehead{1}{l}{b}{$\theta$\tablenote{Zenith angle (degrees).}} & 
	 \tablehead{1}{l}{b}{$z$\tablenote{redshift (when known).}} & 
	 \tablehead{1}{l}{b}{Instrument\tablenote{Instrument reporting the first detection.}} & 
	 \tablehead{1}{l}{b}{$\sigma$\tablenote{Significance of the signal.}} & 
	 \tablehead{1}{l}{b}{UL(fluence)\tablenote{99\% upper limit on the fluence (0.2--20 TeV), in ergs cm$^{-2}$. The numbers in bold take into account absorption by the EBL (using the Primack 05 model). Those with three dots are at redshifts so high that all the emission is expected to be absorbed.}} \\
\hline
041219a & 520   & 26.9  & ... 	& INTEGRAL 	 	& +1.7 & 5.8e-6  \\
050124  & 4    	& 23.0  & ... 	& Swift		  	& -0.8 & 3.0e-7  \\
050319  & 15   	& 45.1  & 3.24 	& Swift		  	& +0.6 & ...  \\
050402  & 8   	& 40.4  & ... 	& Swift		  	& +0.6 & 2.1e-6  \\
050412  & 26   	& 37.2  & ... 	& Swift	 	  	& -0.6 & 1.7e-6  \\
050502  & 20   	& 42.7  & 3.793 & INTEGRAL	 	& +0.6 & ...  \\
050504  & 80   	& 27.6  & ... 	& INTEGRAL	  	& -0.8 & 1.3e-6  \\
050505  & 60   	& 28.9  & 4.3 	& Swift		  	& +1.2 & ...  \\
050509b & 0.128 & 10.0  & 0.226?& Swift	 	  	& -0.9 & \textbf{1.1e-6}  \\
050522  & 15   	& 22.9  & ... 	& INTEGRAL	  	& -0.6 & 5.1e-7  \\
050607  & 26.5  & 29.3  & ... 	& Swift	 	 	& -0.9 & 8.9e-7  \\
050712  & 35   	& 38.8  & ... 	& Swift	 	  	& -0.1 & 2.5e-6  \\
050713b & 30   	& 44.2  & ... 	& Swift	 	  	& -0.3 & 4.0e-6  \\
050715  & 52   	& 36.9  & ... 	& Swift	 	  	& -1.5 & 1.7e-6  \\
050716  & 69   	& 30.3  & ... 	& Swift	 	  	& -0.5 & 1.6e-6  \\
050820  & 20   	& 21.9  & 2.612 & Swift	 	 	& +0.2 & ...  \\
051103  & 0.17  & 49.9\tablenote{This burst was analyzed despite its large zenith angle due to its 
potential interest as a nearby short burst.} 	& 0.001?& IPN		& -0.2 & \textbf{4.2e-6} \\ 
051109  & 36    & 9.7   & 2.346 & Swift		  	& -1.1  & \textbf{4.3e-3}  \\
051111  & 20   	& 43.7  & 1.55 	& Swift		  	& +0.7 & \textbf{3.8e-2}  \\
051211b & 80	& 33.3	& ...	& INTEGRAL		& +0.4	& 2.6e-6	\\

\hline
\end{tabular}
\caption{List of GRB in the field of view of Milagro in the Swift Era (December 2004 -- December 2005), with preliminary upper limits.\label{grb_table}}
\end{table}

\begin{table}
\begin{tabular}{lllllll}
\hline
	 \tablehead{1}{l}{b}{GRB} &
	 \tablehead{1}{l}{b}{Burst Time\tablenote{BAT trigger time (UTC second of day).}} &
	 \tablehead{1}{l}{b}{Flare Time\tablenote{Time of onset of flare (seconds after BAT trigger time).}} &
	 \tablehead{1}{l}{b}{Dur.\tablenote{Approximate duration of flare (s).}} &
	 \tablehead{1}{l}{b}{$\theta$\tablenote{Zenith angle (degrees).}} &
	 \tablehead{1}{l}{b}{$\sigma$\tablenote{Significance of the signal.}} &
	 \tablehead{1}{l}{b}{UL(fluence)\tablenote{99\% upper limit on the fluence (0.2--20 TeV), in ergs cm$^{-2}$. No absorption is taken into account as the redshift of these bursts is unknown.}} \\
\hline
050607  & 33,083 	& 110 &	120 	& 29.1  & -1.3 & 1.5e-6  \\
050607  & 33,083 	& 260 &	340  	& 28.9  & -1.3 & 2.4e-6  \\
050712  & 50,427.5 	& 170 &	265  	& 38.5  & +1.8 & 9.7e-6  \\
050712  & 50,427.5 	& 435 &	255   	& 38.1  & +2.3 & 1.0e-5  \\
050716  & 45,364 	& 135 &	65   	& 30.7  & -1.1 & 1.3e-6  \\
050716  & 45,364 	& 330 &	120   	& 31.4  & -0.4 & 2.3e-6  \\

\hline
\end{tabular}
\caption{Bright X-ray Flares in Swift GRB Afterglows\label{grb_table2}}
\end{table}

\section{Conclusion}

A search for VHE emission from GRBs was performed with the Milagro observatory in the range of 100 GeV
to 100 TeV. A total of 20 satellite-triggered GRBs were well localized and fell within Milagro's field 
of view in the year since the launch of Swift. In addition, six bright X-ray flares from GRB afterglows
were searched for VHE emission. No significant emission was detected from either the bursts or the 
flares. 99\% confidence upper limits on the fluence are presented. 
 
\begin{theacknowledgments}
Many people helped bring Milagro to fruition.  In particular, we
acknowledge the efforts of Scott DeLay, Neil Thompson and Michael Schneider. 
This work has been supported by the 
National Science Foundation (under grants 
PHY-0075326, 
-0096256, 
-0097315, 
-0206656, 
-0245143, 
-0245234, 
-0302000, 
and
ATM-0002744) 
the US Department of Energy (Office of High-Energy Physics and 
Office of Nuclear Physics), Los Alamos National Laboratory, the University of
California, and the Institute of Geophysics and Planetary Physics.
\end{theacknowledgments}


\end{document}